\documentclass[usenat,traditabstract
]{aa}

			\usepackage{amsmath}
			\usepackage{amssymb}
		
			\usepackage{graphicx}
		
			\usepackage{txfonts}
			\usepackage{natbib}
			\bibpunct{(}{)}{;}{a}{}{,} %% natbib format like A&A and ApJ 
			
			\usepackage{url}	% to display ULRs correctly
		
		\newcommand{\aps}{angular power spectrum}
		\newcommand{\apsa}{angular power spectra}

\title{}
\author{}

\begin{document}

	%======================================================================================================================
	% Title + Abstract
	%======================================================================================================================
   
		\title{AGN and QSOs in the eROSITA All-Sky Survey}
		
		% in case of long title for the header of the pages
		\titlerunning{AGN and QSOs in the eRASS}
		
		\subtitle{Part II: The large-scale structure}
		
		\author{Alexander Kolodzig\inst{1}, Marat Gilfanov\inst{1,2}, Gert H\"{u}tsi\inst{1,3}, Rashid Sunyaev\inst{1,2}}
		
		% in case of long author-list for the header of the pages
		\authorrunning{A. Kolodzig et al. (2O!3)}
		
		\institute{%
			Max-Planck-Institut f\"{u}r Astrophysik (MPA), Karl-Schwarzschild-Str. 1, D-85741 Garching, Germany, \email{alex@mpa-garching.mpg.de}
			\and
			Space Research Institute (IKI), Russian Academy of Sciences, Profsoyuznaya ul. 84/32, Moscow, 117997 Russia
			\and
			Tartu Observatory, T\~oravere 61602, Estonia
		}
		
		\date{Received 02.05.2O!3; accepted 22.08.2O!3}
		%!!! The proper receipt and acceptance dates of your manuscript will be set by the editors and inserted by the publisher.

		\abstract 
		% !!! error message for an abstract exceeding 300 words.
		{ 
		The four-year X-ray all-sky survey (eRASS) of the eROSITA telescope aboard the Spektrum-Roentgen-Gamma satellite will detect about $3$~million active galactic nuclei (AGN) with a median redshift of $z\approx1$ and a typical luminosity of $L_{0.5-2.0\,\mathrm{keV}} \sim10^{44}\,\mathrm{erg\;s^{-1}}$.
		We show that this unprecedented AGN sample, complemented with redshift information, will supply us with outstanding opportunities for large-scale structure research.
		For the first time, detailed redshift- and luminosity-resolved studies of the bias factor for X-ray selected AGN will become possible.
		The eRASS AGN sample will not only improve the redshift- and luminosity-resolution of these studies, but will also expand their luminosity range beyond $L_{0.5-2.0\,\mathrm{keV}} \sim10^{44}\,\mathrm{erg\;s^{-1}}$, thus enabling a direct comparison of the clustering properties of luminous X-ray AGN and optical quasars.
		These studies will dramatically improve our understanding of the AGN environment, triggering mechanisms, the growth of supermassive black holes and their co-evolution with dark matter halos.
		
		% Complemented with redshift data,
		The eRASS AGN sample will become a powerful cosmological probe.
		It will enable detecting baryonic acoustic oscillations (BAOs) for the first time with X-ray selected AGN.
		With the data from the entire extragalactic sky, BAO will be detected at a $\ga10\sigma$ confidence level in the full redshift range and  with $\sim8\sigma$ confidence in the $0.8<z<2.0$ range, which is currently not covered by any existing BAO surveys.
		To exploit the full potential of the eRASS  AGN sample, photometric and spectroscopic surveys of large areas and a sufficient depth will be needed.
		}
		
		% A maximum of 6 key words should be listed after the abstract
		% http://www.aanda.org/index2.php?option=com_content&task=view&id=170&Itemid=256
		\keywords{%
			Surveys 
			-- X-rays: general
			-- Quasars: general
			-- Galaxies: active
			-- Cosmology: Large-Scale Structure of Universe
			%-- ?
			}
		
		\maketitle

	%======================================================================================================================
	% Content
	%======================================================================================================================

		%----------------------------------------------------------------------------------------------------------------------
		\section{Introduction} \label{sec:intro}
		%----------------------------------------------------------------------------------------------------------------------
		
%		\subsection{LSS Intro}
		Large-scale structure (LSS) studies are established as an important tool for studies in two major areas of astrophysics: cosmology, and galaxy evolution.
		A key of their success is the increasing number of surveys at different wavelengths with increasing depths and sky coverages.
		In X-rays, many deep, extragalactic surveys have been performed in the past decade \citep{Brandt2005,Cappelluti2012,Krumpe2013}.
		However, in comparison with other wavelengths, X-ray surveys with a large sky coverage and sufficient depth are still rare.
%		\subsection{eROSITA Intro}
		The previous  X-ray all-sky survey was performed by ROSAT%
			\footnote{\url{http://www2011.mpe.mpg.de/xray/wave/rosat/}}
		\citep{Truemper1993,Voges1999} more than two decades ago.
		Its successor with an $\sim30$ times better sensitivity will be the four-year long all-sky survey (eRASS) of the eROSITA\footnote{\url{http://www.mpe.mpg.de/eROSITA}} telescope \citep{eROSITA}, to be launched aboard the Russian Spektrum-Roentgen-Gamma satellite\footnote{\url{http://hea.iki.rssi.ru/SRG}} in 2014. %2014 (any news here?).
		
		The major science goals of the eROSITA mission are studying cosmology with clusters of galaxies and active galactic nuclei (AGN) and constraining the nature of dark matter (DM) and dark energy.
		For a comprehensive description of the eROSITA mission we refer to the science book of eROSITA \citep{eROSITA.SB}. \defcitealias{eROSITA.SB}{SB}
		
%		\subsection{Aim of the paper}
		In this work we explore the potential of studying LSS with the AGN sample to be detected in eRASS.
		We focus on two important aspects of LSS studies: the clustering strength (represented by the linear bias factor, Sect.~\ref{sec:Bias}) and the baryonic acoustic oscillations (BAOs, Sect.~\ref{sec:BAOs}). 
		To measure the former quantity, the redshift accuracy of photometric surveys is sufficient, therefore bias studies can be successfully conducted during and soon after the time eRASS is concluded.
		The BAO measurements, on the other hand, will be much more difficult to accomplish because spectroscopic redshift accuracy over large sky areas will be required.
		Note that a sufficient redshift accuracy can also be provided by high-quality narrow-band multifilter photometric surveys.

		In our previous work \citep{Kolodzig2012}, we have studied the statistical properties of the AGN sample of eRASS and will adopt these results here.
		In the current work, we focus on the AGN detected in the soft-energy band ($0.5-2.0\,\mathrm{keV}$) and on  the extragalactic sky ($|b|>10\degr$, $\sim34\,100\,\mathrm{deg^2}$).
		In the following calculations we assumed the four-year average sensitivity of $1.1\times10^{-14}\;\mathrm{erg\;s^{-1}\,cm^{-2}}$ adopted in \citet[Table~1]{Kolodzig2012}.
		
%		\subsection{Redshift-Information}
		Large optical follow-up surveys will be needed to provide identification and redshift information to the desired accuracy for all eRASS AGN. % (discussed in Sect.~\ref{ssec:z}).
		Current optical surveys are not sufficient in size and/or depth.
		A sensitivity of $I\approx22.5\,\mathrm{mag}$ ($R\approx23.0\,\mathrm{mag}$) is required to detect at least $95\,\%$ of the eRASS AGN \citep{Kolodzig2012}.
		Many photometric and spectroscopic surveys with different parameters have been proposed or are being currently in constructed \citep[e.g.][]{eROSITA.SB}.
		For the purpose of our investigation we assumed that redshifts are available for all eRASS AGN.
		We will explore the effects of redshift-errors in a forthcoming paper \citep{Huetsi2013}.
		
%		\subsection{General Parameters used in the paper}
		We assumed for this work a flat $\Lambda$CDM cosmology with the following parameters:  %gert email 08.03.2012
		$H_0 = 70\,\mathrm{km\,s^{-1}\,Mpc^{-1}}$ ($h=0.70$),
		$\Omega_\mathrm{m} = 0.30$ ($\Omega_\Lambda = 0.70$),
		$\Omega_\mathrm{b} = 0.05$,
		%$\Omega_\mathrm{k} = 0.00$,
		$\sigma_8=0.8$.
		We fixed $H_0$ and $\Omega_\mathrm{m}$ at the values assumed in deriving the  X-ray luminosity functions used in our calcualtions (Sect.~\ref{sec:APS}), 
		the  $\Omega_\mathrm{b}$ and $\sigma_8$ are taken from \citet{Komatsu2011}.
		Luminosities are given for the soft-energy band ($0.5-2.0\,\mathrm{keV}$), and we used the decimal logarithm throughout the paper.

		%----------------------------------------------------------------------------------------------------------------------
		\section{Angular power spectrum} \label{sec:APS}
		%----------------------------------------------------------------------------------------------------------------------
		
			%-----------------------------------------------------------
			% Figure: Angular power spectrum 
			%-----------------------------------------------------------
				\begin{figure} %[htp]
					\resizebox{\hsize}{!}{\includegraphics{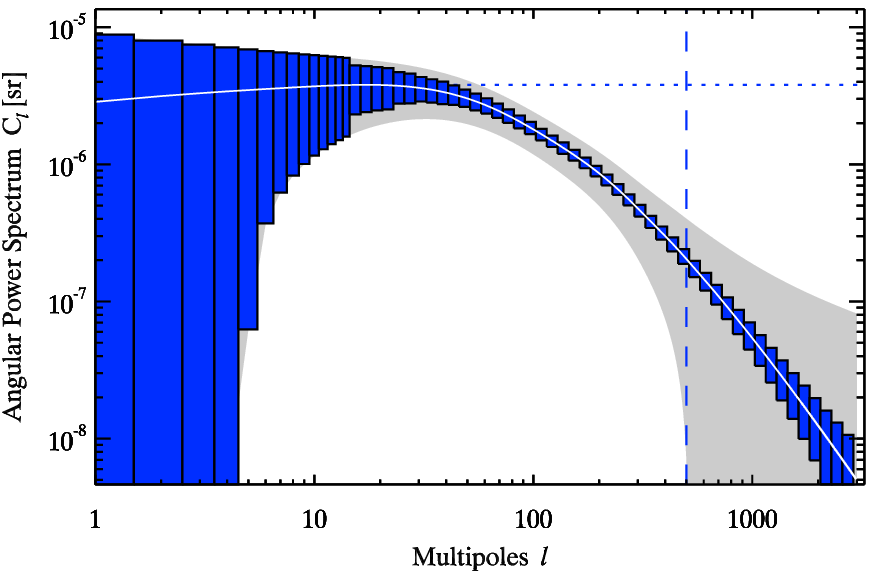}}  
					\caption{\label{fig:APS01_a}%
						Angular power spectrum of the full eRASS AGN sample (soft band, four years) for the extragalactic sky ($f_\mathrm{sky}\approx0.83$) and $0 < z < 5$.
						The grayshaded area and the blue histogram show the $1\sigma$ uncertainty region (Eq.~\ref{eq:Cl_Var01}) without and with $\ell$-binning, respectively.
						The horizontal dotted line shows the level of shot noise, which was already subtracted from the angular power spectrum.
						For multipoles above the vertical dashed line (representing $l_\mathrm{max}\approx500$) our assumption of a linear clustering starts to break down.
						Therefore, we did not consider these multipoles in our subsequent calculations.
					}
					\vspace{5pt}
					\resizebox{\hsize}{!}{\includegraphics{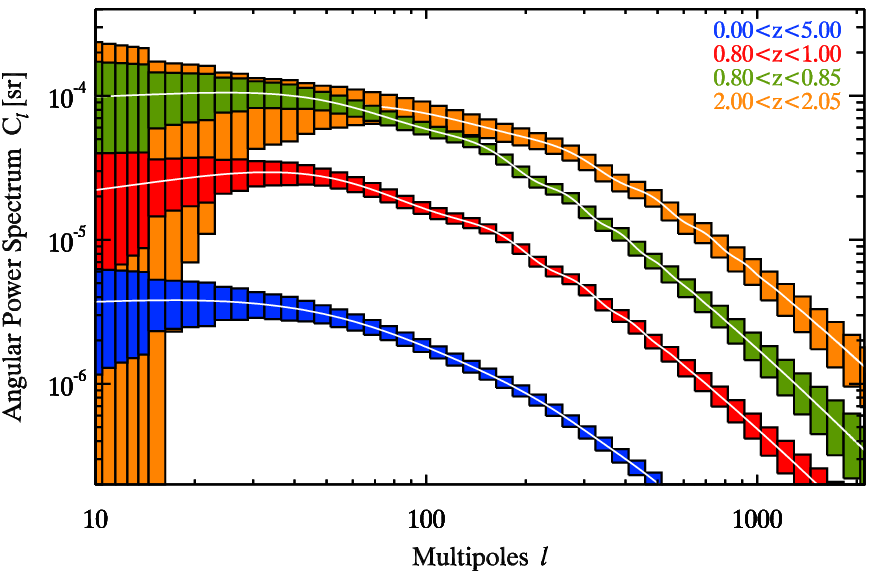}}
					\caption{\label{fig:APS01_c}%
						Same as Fig.~\ref{fig:APS01_a}, but with the angular power spectra for various narrow and broad redshift ranges added (see Sect.~\ref{ssec:APS_res}).
					}
				\end{figure}
			%-----------------------------------------------------------
		
%		\subsection{Intro}
		The commonly used tool for studying LSS is to measure and analyze the clustering of objects (such as AGN) with the 2-point correlation function (2pCF) or the power spectrum (PS) \citep[e.g.,][]{Peebles1980}.
		These two methods, 2pCF and PS, have their benefits and disadvantages \citep[e.g.][]{Wall2012} but contain the same information about the LSS because they are related via Fourier transform.
		We used the \aps{} $C_\ell$ to characterize the clustering properties of objects.
		To predict the power spectra that will be measured with eRASS AGN, we relied on the model for AGN clustering of \citet{Huetsi2012}.
		In particular, we used their model II, where they employed an observationally determined AGN X-ray luminosity function (XLF) and assumed that the linear bias factor of AGN corresponds to the fixed effective mass of the DM halo (DMH).
		The details of our calculations are summarized below.
		
 %		\subsection{Method}
		We calculated the \aps{} as follows:
			\begin{align} \label{eq:Proj}
				C_\ell & = \frac{2}{\pi}\,\int \; P(k) \; \big[W_\ell(k)\big]^2 \; k^2 \, \mathrm{d} k
				\qquad \text{,}
			\end{align}
		where the projection kernel is
			\begin{align} \label{eq:Kern}
				 W_\ell(k) & = \int\limits_{z_\mathrm{min}}^{z_\mathrm{max}} \;f(z)\;g(z)\;b(M_\mathrm{eff},z)\; j_\ell\big(k\,r(z)\big) \;\mathrm{d} z
				 \qquad \text{.}
			\end{align}
		Here,
		$P(k)$ is the 3D linear power spectrum at $z=0$, for which we used the fitting formulae of \citet{Eisenstein1998}, 
		$f(z)$ is the normalized radial selection function, 
		$g(z)$ is the linear growth function \citep[e.g.][]{Dodelson2003}, % Dodelson2003 , p.206
		$b(M_\mathrm{eff},z)$ is the AGN linear clustering bias factor, and
		$j_\ell$ are the spherical Bessel functions on the order of $\ell$,
		where $r(z)$ is the co-moving distance to redshift $z$ \citep[e.g.][]{Hogg1999}.
		
		The radial selection function is defined as the (normalized) differential redshift distribution of AGN,
		which we calculated with the AGN XLF, $\phi(\log L,z)$, of \citet{Hasinger2005}%
		  \footnote{See \citet{Kolodzig2012} for details.}.
		It is the only quantity that contains the information about eRASS, since it depends on the survey sensitivity ($S$) as follows: % of eRASS \citep{Kolodzig2012}: %Eq.(11)
			\begin{align} \label{eq:dNdz}
				\dfrac{\mathrm{d}\mathcal{N}}{\mathrm{d}z}(S,z) & =
					\dfrac{\mathrm{d}V(z)}{\mathrm{d}z} \; \int\limits_{\log L_\mathrm{min}(S,z)}^{\log L_\mathrm{max}} \phi(\log L,z) \,\mathrm{d}\log L
				\qquad \text{.}
			\end{align}
		Here, $\tfrac{\mathrm{d}V(z)}{\mathrm{d}z}$ $[\mathrm{Mpc^3\,sr^{-1}}]$ is the co-moving volume element
		and $L_\mathrm{min}(S,z) = 4\pi\,S\,d_\mathrm{L}^2(z)$, where $d_\mathrm{L}(z)$ is the luminosity distance \citep[e.g.][]{Hogg1999}.

		The AGN linear clustering bias factor, $b(M_\mathrm{eff},z)$, was computed with the analytical model of \citet{Sheth2001} by assuming an effective mass $M_\mathrm{eff}$ of the DMH where the AGN reside.
		Based on recent observations that cover the redshift range to $z\sim3$ \citep[e.g.][]{Allevato2011,Krumpe2012,Mountrichas2013}, we assumed an effective mass of $M_\mathrm{eff}=2\times10^{13}\,h^{-1}\,\mathrm{M_{\sun}}$.
		
		We only focused on the linear clustering regime.
		Therefore,
		we restricted our calculations to spatial co-moving scales larger than $k_\mathrm{max}\approx0.2\,h\,\mathrm{Mpc^{-1}}$, corresponding to wavelengths longer than $\approx30 \,h^{-1}\,\mathrm{Mpc}$.
		% r = 2*pi/k ~ 6/0.2 = 60/2 = 30
		The associated multipole number  is $\ell_\mathrm{max}(\bar{z}) = k_\mathrm{max} \; r(\bar{z})$ and depends on the median redshift $\bar{z}$ of the considered redshift bin.
		At the median redshift of eRASS AGN sample this is $\ell_\mathrm{max}(z\approx1)\sim500$.
		Thus, for our calculations we did not consider $C_\ell$ at multipoles higher than $\ell_\mathrm{max}$.
		
		%RESTORE, FILENAME='~/Projects/2_LSS-Study_eRositaAGN/IDL/functions.sav' eRositaAGN
		%@~/Projects/2_LSS-Study_eRositaAGN/IDL/global_variables.txt
		%r = ComDist(1.0d)/mpc * (H_0/100d) = 3304 Mpc * 0.7 = 2313 Mpc/h
		%l_max = r * 0.2 h/Mpc = 463
		
		For simplicity, we did not take into account linear redshift space distortions (RSD) \citep{Kaiser1987}.
% 		RSD describe the effect that the apparent clustering of galaxies in redshift space is altered by the peculiar velocity of galaxies.
% 		At large-scales, the peculiar velocity caused by the in-fall towards large over-dense regions leads to an apparent contraction of the structure in the line-of-sight, also called the \emph{Kaiser effect} \citep{Kaiser1987}, and to an increase of the clustering signal in redshift space.
		Since the signal-to-noise ratio (S/N) in our \apsa{} is considerably poor at small multipoles (see Figs.~\ref{fig:APS01_a} and~\ref{fig:APS01_c}) where the linear RSD become most significant, we do not expect that our results would change significantly if we would consider them in our calculations.

		\subsection{Uncertainties} \label{ssec:un}
		
		The variance of the $C_\ell$ can be well approximated with
		\begin{align} \label{eq:Cl_Var01} 
			\sigma_{C_\ell}^2 = \frac{2}{(2\ell+1)\;f_\mathrm{sky}} \; \left(  C_\ell + \frac{1}{\mathcal{N}} \right)^2 %\qquad \text{,}
		\end{align}
		assuming Gaussian statistics of the matter fluctuations ($\ell\la\ell_\mathrm{max}$).
		Here, $f_\mathrm{sky}$ is the sky fraction, 
		%with which we can rescale the error to a desired sky coverage,
		which takes into account the effective loss of modes due to partial sky coverage,
		and $\mathcal{N}$ is the AGN surface number density [$\mathrm{sr^{-1}}$], which is computed with the AGN XLF and the survey sensitivity of eRASS \citep[see][]{Kolodzig2012}.
		The first term ($C_\ell$) in the brackets represents the cosmic variance and becomes important at large scales (small $\ell$).
		The second term, the shot noise ($\mathcal{N}^{-1}$), takes into account that we are using a discrete tracer (AGN)
		and becomes dominant at small scales (large $\ell$), where $\mathcal{N}^{-1} >> C_\ell$ (see e.g.\ Fig.~\ref{fig:APS01_a}).
		To minimize the uncertainty in  $C_\ell$, both a high sky coverage and a large number density of objects are needed.

		\subsection{Results} \label{ssec:APS_res}
		
		In Fig.~\ref{fig:APS01_a}, we show the expected \aps{} of the full eRASS AGN sample after four years for the entire extragalactic sky.
		By introducing redshift information (available from other surveys), the \aps{} becomes a relevant tool for LSS studies. % clustering analysis.
		In Fig.~\ref{fig:APS01_c}, we can see how its amplitude  increases with a decreasing size of the redshift bin,
		and oscillations (see Sect.~\ref{sec:BAOs}) in the \aps{} become more prominent as well. % ($100 < \ell < 500$).
		We can see from the two \apsa{} of $0.80<z<0.85$ and $2.00<z<2.05$ (with same redshift bin size) in Fig.~\ref{fig:APS01_c} that the turnover of the spectrum and the positions of the oscillations depend on the redshift.
		The amplitudes are also different because the linear bias factor increases with redshift (see Sect.~\ref{sec:Bias}).
		Because the redshift distribution of eRASS peaks around $z\approx0.8$, the number density around $z\approx0.8$ is much higher than at $z\approx2.0$ and therefore the uncertainty of the \aps{} is smaller for $0.80<z<0.85$ than for $2.00<z<2.05$.

		%----------------------------------------------------------------------------------------------------------------------
		\section{Linear bias factor} \label{sec:Bias}
		%----------------------------------------------------------------------------------------------------------------------
		
%		\subsection{Intro}
		The linear bias factor $b$ is an important parameter for the clustering analysis of AGN.
		It connects the underlying DM distribution with the AGN population.
		Observationally, it has so far been a very challenging task to measure this connection with high accuracy, because of low statistics \citep[e.g.][]{Krumpe2010,Krumpe2012,Miyaji2011,Starikova2011,Allevato2011,Allevato2012,Mountrichas2012,Mountrichas2013}.
		%Seperate ACF and CCF?
		With more accurate observational knowledge of the behavior of the linear bias factor with redshift and luminosity and a comparison with simulations, we will be able to improve our understanding of major questions, such as the nature of the AGN environment, the main triggering mechanisms of AGN activity \citep[e.g.][]{Koutoulidis2013,Fanidakis2013} and how supermassive black holes (SMBH) co-evolve with the DMH over cosmic time \citep[e.g.][]{Alexander2012}.
		
			%-----------------------------------------------------------
			% Figure: Bias Factor: Signal-to-noise ratio of the amplitude
			%-----------------------------------------------------------
				\begin{figure} %[htp]
					\resizebox{\hsize}{!}{\includegraphics{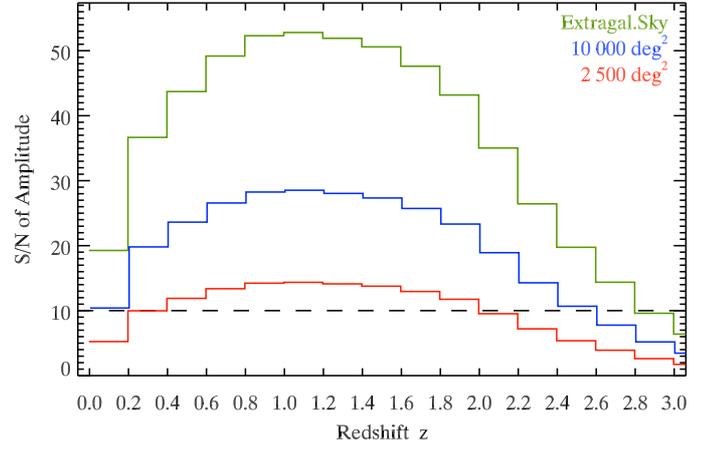}}   
					\caption{\label{fig:SN_Amp_a}%
						Signal-to-noise ratio of the amplitude of the angular power spectrum (Eq.~\ref{eq:SN_bias}) as a function of the redshift for different sky fractions. A $\Delta z=0.2$ binning is assumed.
					}
				\end{figure}
				
				\begin{figure} %[htp]
					\resizebox{\hsize}{!}{\includegraphics{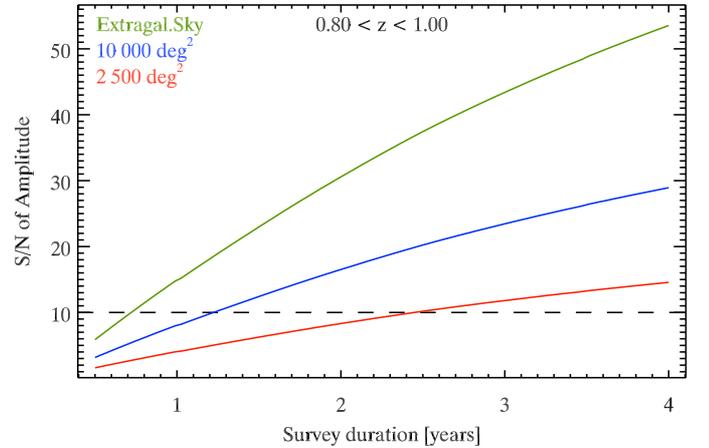}}   
					\caption{\label{fig:SN_Amp_c}%
						Signal-to-noise ratio of the amplitude of the angular power spectrum as a function of the survey duration for the redshift bin $0.8 < z < 1.0$ at different sky fractions.
					}
				\end{figure}
			
				\begin{figure} %[htp]
					\resizebox{\hsize}{!}{\includegraphics{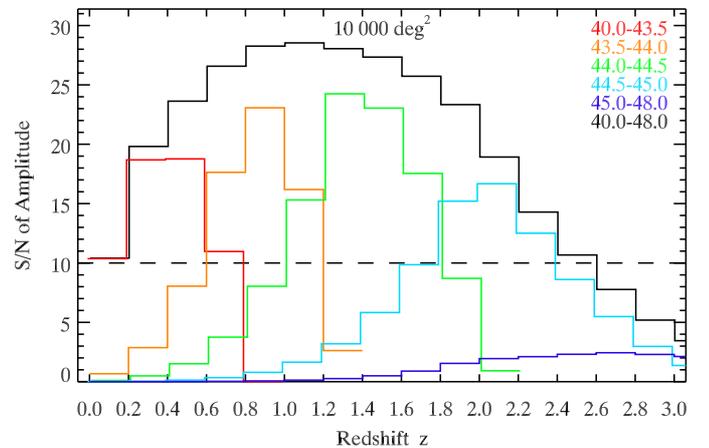}}  
					\caption{\label{fig:SN_Amp_b}%
						Same as Fig.~\ref{fig:SN_Amp_a}, but for different luminosity ranges (in units of $\log(L[\mathrm{erg\,s^{-1}}])$) and a sky coverage of $10\,000\,\mathrm{deg^2}$.
					}
				\end{figure}
			%-----------------------------------------------------------

		\subsection{Method}
		
		The linear bias factor was measured by comparing the amplitudes of the observed PS of tracer objects and of the theoretical PS of the DM, under the assumptions of a certain cosmology.
		Since the PS amplitude of tracer objects is proportional to the square of the linear bias factor ($\propto b^2$), its uncertainty directly reflects the uncertainty of measuring the latter.
		Knowing the amplitude ($A$) of our \aps{}, we are able to estimate this uncertainty.
		The S/N for measuring the normalization of the power spectrum $C_\ell$ assuming that its shape is known is given by
			\begin{align} \label{eq:SN_bias} 
				\frac{S}{N} = \frac{A}{\delta A} = \sqrt{ \sum_{\ell=1}^{\ell_\mathrm{max}(\bar{z})} \left( \frac{C_\ell}{\delta C_\ell} \right)^2 }
				\qquad \text{.}
			\end{align}
 		Here, we assumed that all multipoles are independent.
		
		We used a redshift binning of $\Delta z = 0.20$ for our calculation (see Fig.~\ref{fig:SN_Amp_a} and \ref{fig:SN_Amp_b}) but other bin sizes would also be possible to demonstrate our results.
		In current observations \citep[e.g.][]{Allevato2011,Starikova2011,Koutoulidis2013} typically a much larger bin size is used to achieve a reasonable S/N ratio for $b$ in  each redshift bin. %Seperate ACF and CCF?

		\subsection{Results}
		
		In Fig.~\ref{fig:SN_Amp_a} the achievable S/N of the power spectrum amplitude is shown as a function of redshift.
		The shape of the curves is dominated by the redshift distribution of AGN modified by the quadratic-like increase of the linear bias factor with redshift at constant DMH mass \citep[e.g.][]{Sheth2001}.
		We we are able to measure the amplitude to a high accuracy ($<10\,\%$) for a wide redshift range even with a fairly small fraction of the sky (e.g.\ $\sim2\,500\,\mathrm{deg^2}$).
		
		The analysis of the linear bias factor can be performed before the entire four-year long eRASS is completed, as we can see from Fig.~\ref{fig:SN_Amp_c}.
		For an SDSS-like sky coverage of $10\,000\,\mathrm{deg^2}$ (blue curve) one can work with the data of eRASS after only 1.5~years (three full sky scans) to study the evolution of the linear bias factor to an accuracy of better than $10\,\%$ in the amplitude  for the redshift bin $0.8 < z < 1.0$.
		For a sky region of $2\,500\,\mathrm{deg^2}$ it needs five full sky scans (2.5~years).
		For the neighboring redshift bins $0.6 < z < 0.8$ and $1.0 < z < 1.2$ the results are similar.
		The sensitivities used for this calculation are taken from Fig.~1 of \citet{Kolodzig2012}.
		
		Owing to the high S/N of the power spectrum amplitude, we will be able to separate the AGN into different luminosity groups. 
		This is demonstrated in Fig.~\ref{fig:SN_Amp_b} for a sky coverage of $10\,000\,\mathrm{deg^2}$. 
		We will be able to achieve an accuracy of $<10\,\%$ for most luminosity groups for a certain redshift range.
		This means that it will be possible to perform a redshift- and luminosity-resolved analysis of the linear bias factor of AGN with eRASS with high statistical accuracy.
		We note that in our calculation the difference in the S/N of the luminosity groups in Fig.~\ref{fig:SN_Amp_b} is driven only by the difference in the redshift distribution of eRASS AGN and the redshift dependence of the linear bias factor.

		%----------------------------------------------------------------------------------------------------------------------
		\section{Baryonic acoustic oscillations} \label{sec:BAOs}
		%----------------------------------------------------------------------------------------------------------------------

%		\subsection{Intro}
		Acoustic peaks in the power spectra of matter and CMB radiation are among the main probes for measuring the kinematics of the Universe \citep[e.g.][]{Weinberg2012}.
		They were predicted theoretically more than four decades ago \citep{Sunyaev1970,Peebles1970} and now have become a standard tool of observational cosmology. 
		Unlike acoustic peaks in the angular power spectrum of CMB,  their amplitude  in the matter power spectrum  in the $\Lambda$CDM Universe is small.
		For this reason,  galaxy surveys have only recently reached sufficient breadth and depth for the first convincing detection of BAO, achieved with the SDSS data \citep{Cole2005,Eisenstein2005,Huetsi2006,Tegmark2006}.
		Since then, BAO have been measured  extensively up to redshift $z\sim0.8$, in particular with luminous red giant galaxies (LRGs) \citep[e.g.][]{Anderson2012}.		
		Above this redshift limit, BAO features were only found in the correlation function of the transmitted flux fraction in the Lyman-$\alpha$ forest of high-redshift quasars \citep{Busca2012,Slosar2013}, but have not yet been directly detected in galaxy distribution.
		
		For the as yet uncharted redshift range  from $z\sim0.8$ up to $\sim2.0$, AGN, quasars and emission-line galaxies (ELGs) are proposed to be the best tracers for measuring BAOs, however, currently existing  surveys do not achieve the required statistics for a proper detection \citep{Sawangwit2012,Comparat2013}.
		eRASS and the proposed SDSS-IV%
			\footnote{\url{http://www.sdss3.org/future/}}
		survey program eBOSS%
			\footnote{\url{http://lamwws.oamp.fr/cosmowiki/Project_eBoss}}
		(2014-2020) will be the first surveys
		to change this situation.
		eRASS will achieve a sufficiently high density of objects $\mathcal{N}\sim40\,\mathrm{deg^{-2}}$ in this redshift range and will have by far the largest sky coverage compared with to eBOSS and all other dedicated BAO surveys.
		This would enable one to push the redshift limit of BAO detections in the power spectra of  galaxies far beyond the present limit of $z\sim0.8$.

			%-----------------------------------------------------------
			% Figure: BAOs: P(k)/P_{Smooth}(k)
			%-----------------------------------------------------------
				\begin{figure} %[htp]
					\resizebox{\hsize}{!}{\includegraphics{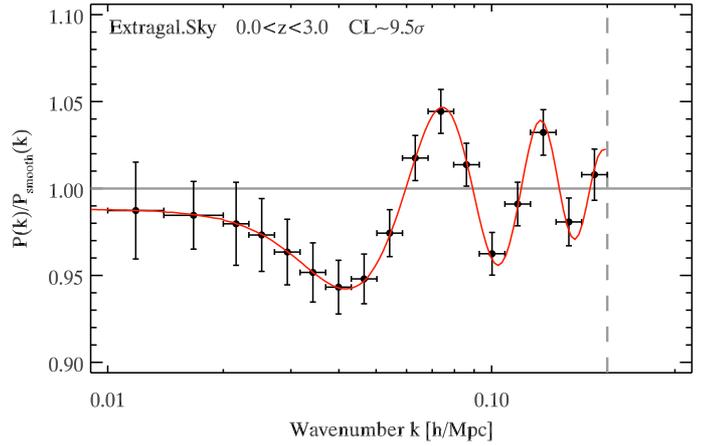}}   
					\caption{\label{fig:BAOs_Ex_005}%
						Baryonic acoustic oscillations in the power spectrum for the extragalactic sky in the redshift range $0.0 < z < 3.0$.
						% with a redshift binning of $\Delta z = 0.05$.
						At wavenumbers above the vertical dashed line (corresponding to $0.2\,h\,\mathrm{Mpc^{-1}}$) our assumption of a linear clustering starts to break down.
						The red curve shows the original input model for the BAOs.
					}
				\end{figure}

		\subsection{Method} \label{ssec:BAO.method}

By construction of the model, BAOs are included in the AGN clustering model of \citet{Huetsi2012} through the  3D linear power spectrum (Sect.~\ref{sec:APS}).
Oscillations can be noticed, for example, in Fig.~\ref{fig:APS01_c} in the power spectra of objects selected in narrow-redshift intervals.
Because the angular scales of acoustic peaks  depend on the redshift, BAO are smoothed out in the power spectra computed for broad-redshift intervals through the superposition of signals coming from many different redshift slices

Although the real data will be analyzed in a much more elaborate way, for the purpose of this calculation we used a simple method to estimate the amplitude and statistical significance of the BAO signal detection.
We divided a broad redshift interval into narrow slices of width $\Delta z$ and for each slice computed the angular power spectrum, $C_\ell(z)$, and converted the multipole number to the wavenumber $k=\ell/r(z)$ to obtain $P(k,z)$.
These power spectra were co-added in the wavenumber space to obtain the total power spectrum $P(k)$ of objects in the broad-redshift interval.
This power spectrum will have unsmeared  BAO features.
To estimate their statistical significance, we also constructed a model $C_{\ell,\,\mathrm{smooth}}(z)$  without acoustic peaks by smoothing the matter transfer function, similar to \citet{Eisenstein1998}.
From this model we computed the smoothed power spectrum in the wavenumber space, $P_{\rm smooth}(k)$, that does not contain BAOs.
To illustrate the amplitude of the BAO signal, one can plot  the difference $P(k)-P_{\rm smooth}(k)$ or the ratio $P(k)/P_{\rm smooth}(k)$.

By analogy with Eq.~\eqref{eq:SN_bias}, the S/N of the BAO detection in the eRASS AGN sample can be computed as
\begin{equation}
	\frac{S}{N} = \sqrt{\sum_z\sum_\ell^{\ell_\mathrm{max}(z)} \left(\frac{C_\ell(z)-C_{\ell,\rm smooth}(z)}{\sigma_{C_\ell}}\right)^2}
	\qquad \text{,}
\end{equation}
where the outer summation was performed over the redshift slices and the variance $\sigma_{C_{\ell}}^2$ was calculated from Eq.~\ref{eq:Cl_Var01}. 

The result of this calculation depends on the choice of the thickness of the  redshift slice $\Delta z$ (see Fig.~\ref{fig:BAO_DeltaZ}).
For too high values of $\Delta z$, BAO will be smeared out, as discussed above (cf. Fig.~\ref{fig:APS01_c}).
On the other hand, for too low values of $\Delta z$, at which the thickness of the redshift slice becomes somewhat thinner than the co-moving linear scale of the acoustic oscillations, the cross-spectra between different redshift slices will need to be taken into account in computing $P(k)$.
For the purpose of these calculations we chose $\Delta z = 0.05$.
The corresponding thickness of the redshift slice at $z\sim 1$ is approximately equal to the co-moving linear scale of the first BAO peak.
Note that omitting of the cross-spectra in our calculation leads to a slight underestimation of the confidence level of BAO detection.

% 		direct calculation of the CL of the BAOs (no k-binning):
% 			\begin{align} 
% 				F_\ell(\bar{z}) & = \frac{C_{\ell,\mathrm{BAOs}}(\bar{z})}{C_{\ell,\mathrm{smooth}}(\bar{z})} \notag \\
% 				\sigma_{F_\ell(\bar{z})} & = \frac{\sigma_{C_{\ell,\mathrm{BAOs}}(\bar{z})}}{C_{\ell,\mathrm{smooth}}(\bar{z})} \notag \\
%  				\frac{S}{N} & = \sqrt{ \sum_{i} \left\{  \sum_{\ell=1}^{\ell_\mathrm{max}(\bar{z}_i)} \left( \frac{F_\ell(\bar{z}_i) - 1}{\sigma_{F_\ell(\bar{z}_i)}} \right)^2 \right\} } \notag
% 			\end{align}

		\subsection{Results}
\begin{figure} %[htp]
					\resizebox{\hsize}{!}{\includegraphics{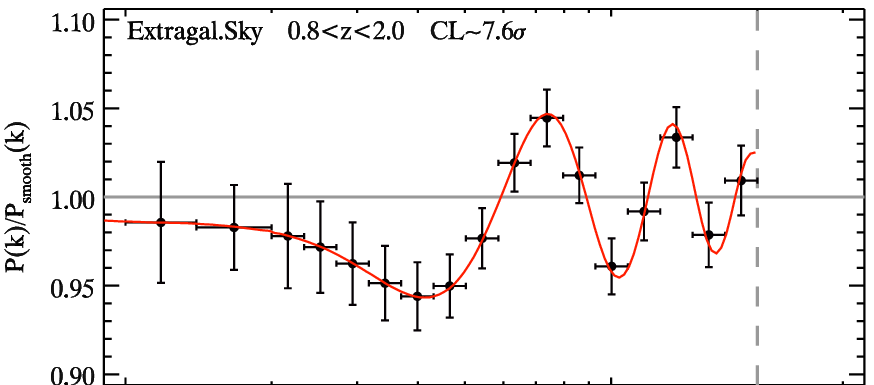}}
					\resizebox{\hsize}{!}{\includegraphics{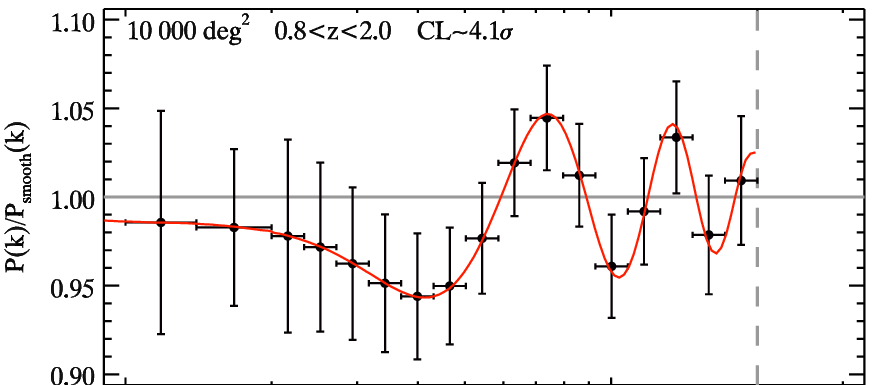}}   
					\resizebox{\hsize}{!}{\includegraphics{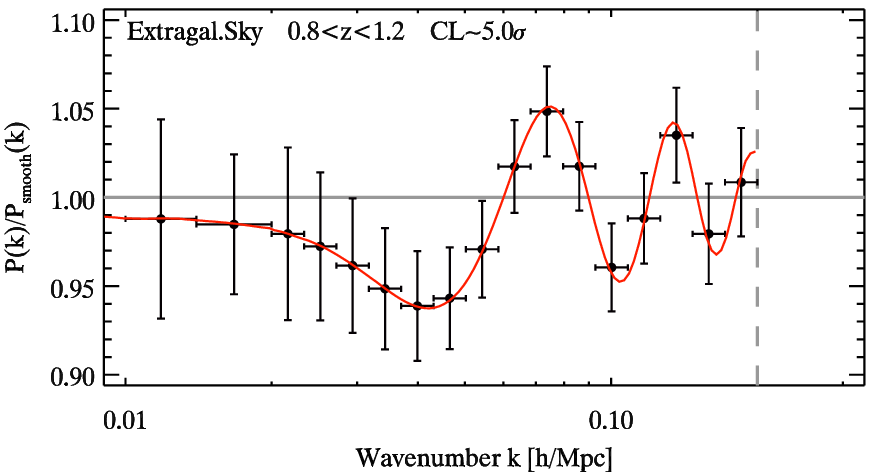}}   
					\caption{\label{fig:BAOs_others}%
						Same as Fig.~\ref{fig:BAOs_Ex_005}, but for different redshift ranges and sky coverages.
					}
				\end{figure}

\begin{figure} %[htp]
					\resizebox{\hsize}{!}{\includegraphics{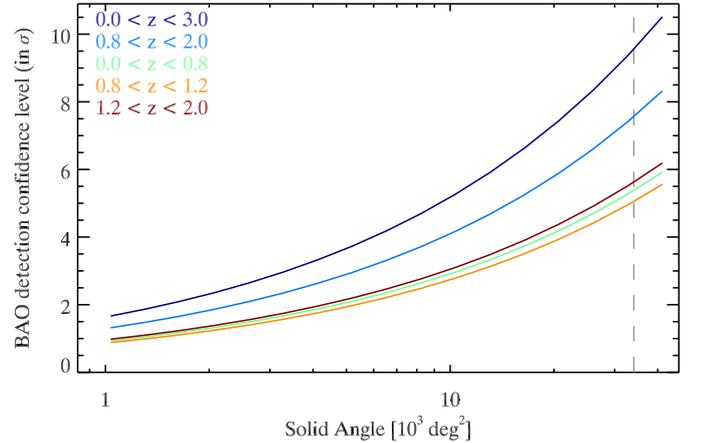}}   
					\caption{\label{fig:BAOs_Sigma}%
						Confidence level of a BAO detection
						%with \apsa{} and a redshift binning of $\Delta z=0.05$
						as a function of sky coverage for different redshift ranges (see Sect.~\ref{ssec:BAO.method} for more explanations).
						The vertical gray dashed line shows the area of the extragalactic sky.
					}
				\end{figure}

	In  Figs.~\ref{fig:BAOs_Ex_005} and  \ref{fig:BAOs_others} we show the ratio $P(k)/P_{\rm smooth}(k)$ along with its uncertainties computed as described above. 
		As these plots demonstrate, with the whole eRASS AGN sample for the extragalactic sky we expect to be able to detect the BAOs with a confidence level (CL) of $\sim10\sigma$ (Fig.~\ref{fig:BAOs_Ex_005}).
		For the currently unexplored  redshift range of $0.8 - 2.0$ a confidence level of $\sim8\sigma$ will be achieved, which can be seen in the top panel of Fig.~\ref{fig:BAOs_others}.
		Decreasing the sky area to $20\,000\,\mathrm{deg^2}$ or $10\,000\,\mathrm{deg^2}$ (see middle panel of Fig.~\ref{fig:BAOs_others}), we obtain $\sim6\sigma$ and $\sim4\sigma$, respectively.
		In Fig.~\ref{fig:BAOs_Sigma} we show that the confidence level of the BAO detection for different redshift ranges depends on the sky coverage.
		The curves follow a $f_\mathrm{sky}^{-0.5}$- dependence, as expected from Eq.~\eqref{eq:Cl_Var01}.
		
		As Fig.~\ref{fig:BAOs_Sigma} shows,  for the redshift ranges  $0.0-0.8$, $0.8-1.2$ and $1.2-2.0$ the achievable confidence levels are very similar, therefore  the power spectra ratio shown in the bottom panel of Fig.~\ref{fig:BAOs_others} is representative for all three redshift intervals.
		Comparing the upper and bottom panels of Fig.~\ref{fig:BAOs_others}, one can see that the BAO signal depends on the redshift range, while the top and middle panels show the degradation due to the reduced sky coverage.

			%-----------------------------------------------------------
		
			%-----------------------------------------------------------
			% Figure: BAOs: detection significance
			%-----------------------------------------------------------
			%-----------------------------------------------------------

		%----------------------------------------------------------------------------------------------------------------------
		\section{Discussion and conclusions} \label{sec:Dis}
		%----------------------------------------------------------------------------------------------------------------------
		
% 		\subsection{Intro?}
% 		\dots

		\subsection{Linear bias factor}

		Measuring of the linear bias factor provides a simple and direct method for estimating the average mass of DMHs that host a given subpopulation of AGN.
		With the eRASS, these measurements will become possible to unprecedented detail.
		The dramatic improvement of the redshift- and luminosity resolution of DMH mass measurements will have a great impact on our understanding of the environment of AGN, AGN triggering mechanisms, and SMBH co-evolution with the DMH.
		
		Observational results of AGN clustering studies suggest a higher DMH mass for AGN than for quasars and a weak dependence between DMH mass and AGN luminosity \citep[e.g.][]{Krumpe2010,Krumpe2012,Krumpe2013,Miyaji2011,Allevato2011,Allevato2012,Cappelluti2012,Mountrichas2012,Mountrichas2013,Koutoulidis2013,Fanidakis2013}.
		However, uncertainties are still large and AGN luminosities available for these studies are typically limited by $L \sim 10^{44}\,\mathrm{erg\;s^{-1}}$ .
		For instance, \citet{Koutoulidis2013} compared their results from clustering studies of  AGN in four  extragalactic X-ray surveys of different depth and coverage (CDFN, CDFS, COSMOS and AEGIS) with the theoretical predictions of \citet{Fanidakis2012}.
		Their goal was to determine the dominant SMBH growth mode for AGN of different luminosities from the AGN bias factor measurements -- either through galaxy mergers and/or disk instabilities  or through accretion of hot gas from the galaxy halo.
		However, the uncertainties of bias factor measurements and of the DMH mass estimates were too large to clearly distinguish the dominant growth mode as a function of luminosity.
		In particular, the luminosity range of objects available for their analysis,  $L \sim 10^{42-44}\,\mathrm{erg\;s^{-1}}$, was too narrow to challenge the prediction of \citet{Fanidakis2012} that luminous galaxies with $L > 10^{44}\,\mathrm{erg\;s^{-1}}$ reside in DMHs of moderate mass of $\sim10^{12}\,M_{\sun}$.
		For the same reason, no direct comparison was possible with the results of the  optical quasar surveys \citep{Alexander2012}.
		To overcome this limitation, \citet{Allevato2011} studied  broad-line (BL) AGN from the COSMOS survey and found for their luminosity bin $L\sim10^{43-46}\,\mathrm{erg\;s^{-1}}$, a significantly higher DMH mass than inferred from quasar studies, suggesting that for broad-line AGN major merger may not be the dominant triggering mechanism, which agrees reasonably well with recent simulations \citep{Draper2012,Hirschmann2012,Fanidakis2012,Fanidakis2013}.
		However, the large width of the luminosity bin required to accumulate sufficient statistics did not allow them to draw a firm conclusion. 
		
		As illustrated by Fig.~\ref{fig:SN_Amp_b}, the eRASS AGN sample will not only dramatically improve the statistics, but will also expand the luminosity range beyond $L\sim10^{44}\,\mathrm{erg\;s^{-1}}$ to the luminosity domain characteristic of quasars.
		Thus, the  eRASS data will not only increase the redshift- and luminosity resolution of DMH mass estimate of AGN, but will open possibilities for a detailed comparison of the clustering properties of luminous AGN and optical quasars.
		Another aspect of bias  measurements with eRASS that determines their uniqueness  is that they are based on the X-ray selected AGN sample and will cover a very broad SMBH mass range, broader  than that in  AGN samples produced by optical/IR or radio surveys \citep[e.g.][]{Hickox2009}.

		The growth rate of SMBHs over time can be measured from the XLF of AGN \citep[e.g.][]{Aird2010} and eRASS will improve the accuracy and redshift resolution of these studies tremendously \citep[e.g.][]{Kolodzig2012}.
		Combined with clustering bias data, these measurements will be placed in a broader context and be connected with DMH properties, which will provide new insights into the co-evolution of SMBHs with their DMHs \citep[e.g.][]{Alexander2012} and will also help to investigate the dominant triggering mechanisms of AGN activity \citep[e.g.][]{Koutoulidis2013,Fanidakis2013}.

The AGN clustering model used here and the corresponding calculations of the AGN linear bias factor ignored  the internal structure of DMHs, that is they were restricted to scales larger than the size of a typical DMH.
Expressed in the language of the halo occupation distribution (HOD) formalism, these calculations  operated with population-averaged halo occupation numbers.
The angular resolution of the eROSITA telescope, $\approx30\arcsec$ FOV averaged HEW, is sufficient to resolve subhalo linear scales.
Clustering measurements on small scales will permit one to obtain a detailed picture of the way AGN are distributed within a DMH (e.g. to measure fractions of central and satellite AGN), as well as how the HOD depend on the DMH mass and redshift, and AGN luminosity \citep{Miyaji2011,Allevato2011,Starikova2011,Krumpe2012}.
		Extrapolating the results of XMM-COSMOS data analysis by \citet{Richardson2013}, we may expect that high-accuracy determination of the HOD parameters will be easily achieved with eRASS data, which will be able to address all these  questions, advancing our understanding of AGN clustering on small scales and their HOD.

		\subsection{BAOs} \label{ssec:DisBAO}
		The BAO detection beyond redshift $\sim0.8$ will be a very significant milestone for the direct measurement of the kinematics of the Universe.
		eRASS will be able to map this uncharted redshift region up to $z\sim2$ with a sufficiently high AGN number density to measure BAOs with a high statistical significance (see Fig.~\ref{fig:BAOs_Sigma}).
		For a proper prediction of the way in which these measurements will improve our constraints on cosmological parameters, Markov chain Monte Carlo (MCMC) simulations \citep[e.g.][]{Lewis2002} and/or Fisher matrix calculations  \citep[e.g.][]{Tegmark1997} have to be made, which is beyond the focus of our work.
		\citet{Sawangwit2012} have performed an MCMC simulation for quasars/quasi-stellar objects (QSOs) and demonstrated that a $3-4\sigma$ BAO detection (of a $3\,000\,\mathrm{deg^2}$ QSO survey with $\mathcal{N}=80\,\mathrm{deg^{-2}}$) for $1.0<z<2.2$ can significantly reduce the uncertainties.
		Although the survey parameters of eRASS differ (much larger sky coverage but smaller source density for the same redshift region,  $\sim40\,\mathrm{deg^{-2}}$), 
		the results of \citet{Sawangwit2012} can give one an idea of how the eRASS AGN sample will  improve the accuracy  of cosmological parameter determination.

		Our calculationd were limited to the linear regime and did not take nonlinear structure growth into account, which would smear out the BAO signal to some extent.
		This would lead to a decreased detection significance \citep[e.g.][]{Eisenstein2007}.
		However, with BAO reconstruction methods one will be able to correct for this effect to some extent \citep[e.g.][]{Padmanabhan2012,Anderson2012}.
		We also note that our confidence level estimates of the BAO detection are fairly conservative because they neglect information contained  in the cross-spectra.
		This will counterbalance the negative effect of BAO smearing, as will be demonstrated in a forthcoming paper \citep{Huetsi2013}, where we will study BAO predictions in a broader context and account for these effects more accurately.
		
		%-----------------------------------------------------------
		% EFFECTIV VOLUME
		\subsubsection{Comparison with dedicated BAO surveys} \label{sss:Veff}
		%-----------------------------------------------------------
		
			%-----------------------------------------------------------
			% Figure: effectiv Volume of d
			%-----------------------------------------------------------
				\begin{figure} %[htp]
					\resizebox{\hsize}{!}{\includegraphics{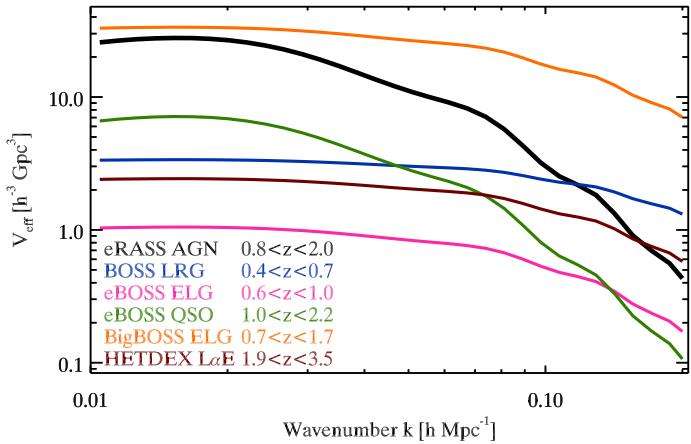}}   
					\caption{\label{fig:Veff}%
						Effective volumes of BAO surveys listed in  Table~\ref{tab:BAO_Surveys} as a function of the wavenumber. Effective volumes are computed for redshift ranges indicated in the plot.
						%The shaded areas around the curves show the $\pm30\,\%$-area, approximately indicating the accuracy of these calculations.
					}
				\end{figure}
			%-----------------------------------------------------------

			%-----------------------------------------------------------
			% Tables: BAO -surveys
			%-----------------------------------------------------------
			%p13a_EffVol_01.pro
		
			\begin{table*}[t]
				\caption{Parameters of BAO surveys.}
				\label{tab:BAO_Surveys}
				\renewcommand{\arraystretch}{1.3} 
				\centering % used for centering Table
				\begin{tabular}{l |  c  c  c  c  c  c  c  c  c} 
\hline\hline 	% inserts double horizontal lines
Survey	& Tracer	& $b_\mathrm{tr}$	& Redshift	& $\Omega_\mathrm{survey}$	& $\mathcal{N}$	& $\left<n\right>$				& $V_\mathrm{eff}$ [h$^{-3}$ Gpc$^3$]
			& Implem.	& Ref.	\\
	& Object	& $(z=0)$		& Range		& [$10^3\,$deg$^2$]		& [deg$^{-2}$]	& [10$^{-4}\,$h$^3$ Mpc$^{-3}$]	& $(k=0.07\,h\,\mathrm{Mpc^{-1}})$
			& Date	\\
\hline
eRASS	& AGN		& $1.33$		& $0.8\!-\!2.0$	& $34.1$	& $40$		& $0.12$	& $7.8$	& $2014\!-\!2018$	& \\
BOSS	& LRG		& $2.00$		& $0.4\!-\!0.7$	& $10.0$	& $80$		& $2.3$		& $2.9$	& concluded		& [1]	\\
eBOSS	& ELG		& $1.00$		& $0.6\!-\!1.0$	& $1.5$		& $180$		& $2.1$		& $0.5$	& $2014\!-\!2020$	& [2]	\\
eBOSS	& QSO		& $1.20$		& $1.0\!-\!2.2$	& $7.5$		& $90$ 		& $0.21$	& $1.7$	& $2014\!-\!2020$	& [2]	\\
BigBOSS	& ELG		& $0.84$		& $0.7\!-\!1.7$	& $14.0$	& $1730$	& $6.3$		& $24.0$	& $>2020$	& [3]	\\
HETDEX	& L$\alpha$E	& $2.34$ \tablefootmark{(a)}	& $1.9\!-\!3.5$	& $0.45$	& $3370$ & $7.2$	& $1.9$	& $2014\!-\!2017$	& [4]	\\
\hline %inserts single line
				\end{tabular}
				\tablefoot{
					\tablefoottext{a}{for $z\approx2.2$} \\
					Tracer object: AGN - active galactic nuclei, LRG - luminous red giant galaxy, ELG - emission line galaxy, QSO - quasar/quasi-stellar object, L$\alpha$E - Lyman-$\alpha$ emitting galaxy \\
					Columns: $b_\mathrm{tr}$ - bias factor of the tracer object;
					$\Omega_\mathrm{survey}$ - solid angle covered by the survey;
					$\mathcal{N}$ - surface number density;
					$\left<n\right>$ - average volume number density;
					$V_\mathrm{eff}$ - effective survey volume at the first BAO peak \\
					References:
					[1] \citet{Anderson2012} and \citet{Dawson2013};
					[2] eBOSS team, priv. comm.;
					[3] \citet{BigBOSS};
					[4] HETDEX team, priv. comm.
					%Method: A) with $\mathrm{d}n/\mathrm{d}z$ and $b(z)$ ;  B) with $\mathrm{d}n/\mathrm{d}z$ and $b_{z=0}$ ; C) with $\bar{n}$ and $b_{z=0}$
				}
			\end{table*}
			%-----------------------------------------------------------
			
		We now compare the potential of the eRASS AGN sample  with dedicated BAO surveys in the optical band.
		For the latter, we consider the completed BOSS CMASS survey \citep{Anderson2012}, the planned eBOSS%
			\footnote{\url{http://lamwws.oamp.fr/cosmowiki/Project_eBoss}}
		and HETDEX \citep{HETDEX} surveys that are scheduled to start in 2014, and the future BigBOSS survey \citep{BigBOSS}, anticipated to be operrational in 2020 time.
		Table~\ref{tab:BAO_Surveys} summarizes the key parameters of these surveys that are relevant for the BAO studies.
		
		A quantity often used to estimate the statistical performance of a galaxy clustering survey is its effective volume \citep[e.g.][]{Eisenstein2005}:
			\begin{align} \label{eq:Veff}
				V_\mathrm{eff}(k) & = \Omega_{\rm survey}
					  \int\limits_{z_\mathrm{min}}^{z_\mathrm{max}} \left[ \dfrac{n(z) \,P_\mathrm{tr}(k,z)}{n(z)\,P_\mathrm{tr}(k,z) + 1} \right]^2 \dfrac{\mathrm{d}V(z)}{\mathrm{d}z\,\mathrm{d}\Omega} \,\mathrm{d}z
				\qquad \text{,}
			\end{align}
		where $\Omega_{\rm survey}$ is the solid angle covered by the survey, $P_\mathrm{tr}(k,z)= b_\mathrm{tr}^2(z)[g(z)/g(0)]^2 P(k)$ is the power spectrum of objects used as LSS tracer, $b_\mathrm{tr}(z)$ is their redshift-dependent bias factor
		and $n(z)$ is their redshift distribution [h$^3$ Mpc$^{-3}$].
		Other quantities are defined in the context of Eqs.~\eqref{eq:Proj}--\eqref{eq:dNdz}.
		For optical surveys we assumed that $b_\mathrm{tr}(z)g(z)=\mathrm{constant}$, therefore we only need to compute $P_\mathrm{tr}(k,z=0)$.
		For HETDEX we used $P_\mathrm{tr}(k,z=2.2)$, because the linear bias factor of HETDEX tracer objects was estimated by comparing the power spectrum of Lyman-$\alpha$ emitting galaxies (L$\alpha$Es) from the simulations of \citet{Jeong2009} at $z\approx2.2$ with our linear (DM) power spectrum $P(k)$ transformed to $z=2.2$.
		The $n(z)$ dependencies for optical surveys were taken from references listed in Table~\ref{tab:BAO_Surveys}.
		
		The results of these calculations are plotted in  Fig.~\ref{fig:Veff} where we show the effective volumes of different surveys as a function of the wavenumber.
		Their values at the first BAO peak are listed in the respective column of Table~\ref{tab:BAO_Surveys}.
		In these calculations the integration in Eq.~\eqref{eq:Veff} was performed over the best-fit redshift range of each survey, as listed in Table~\ref{tab:BAO_Surveys}.
		For eRASS we used the $z=0.8-2.0$ range to emphasize its strength in this uncharted redshift region.
		The eRASS effective volume for the full redshift range is $\sim70\%$ higher for the first BAO peak.
		 
The result of the effective volume calculations obviously depends on the assumptions made of values and redshift dependences of co-moving density, bias,  and the growth factor, which are not always precisely known, especially for the future surveys.
Furthermore, efficiencies  of redshift determinations are expected to be between $50\,\%$ and $80\,\%$, but their exact values are difficult to predict.
To have a fair comparison, we assumed a $100\,\%$ efficiency for all future surveys.
% To approximately indicate the limited accuracy of these calculations, we show in Fig.~\ref{fig:Veff}  the $\pm 30\%$ area around the curves by shaded regions \citep[cf.][]{Eisenstein2005}.
Nevertheless, these curves should give a reasonably accurate comparison of the different surveys' qualities in measuring the power spectrum at different scales (note that the uncertainty of the power spectrum  is proportional to $V_\mathrm{eff}^{-0.5}$).
		
		We can see from Fig.~\ref{fig:Veff} that the effective volumes of  eRASS AGN and eBOSS QSO samples fall more rapidly towards smaller scales than for other surveys.
		This is a consequence of the lower volume density of X-ray selected AGN and optical QSOs (Table~\ref{tab:BAO_Surveys}).
		For the same reason, the statistical errors in the eRASS AGN power spectrum  are dominated by the shot noise, but the high sky coverage of eRASS keeps them small.
		As one can see from the figure,  eRASS is more competitive at larger scales, up to the second and third BAO peaks, where its sensitivity becomes similar to BOSS and HETDEX, respectively.
		It should be noted, however, that BOSS and HETDEX cover a relatively low ($z\la 0.7$) and high ($z\ga2$) redshift domain, respectively, whereas all other surveys presented in Fig.~\ref{fig:Veff} are aimed for the redshift region between $\sim0.8$ and $\sim2.0$ (Table~\ref{tab:BAO_Surveys}).
		Around the first peak the effective volume of eRASS is a factor of about $2-4$ higher than for BOSS and HETDEX, but a few times lower than that of BigBOSS (Table~\ref{tab:BAO_Surveys}).
		On the other hand, eRASS exceeds eBOSS at all wavenumbers.
		This would still be the case when one were to consider the subset of the eRASS sample to cover only $\sim1/3$ of the extragalactic sky.
		
		In conclusion we note that it is remarkable that the statistical strength of eRASS for BAO studies is similar to that of dedicated BAO surveys, even though  eRASS was never designed for this purpose.
		Potentially, the eRASS AGN sample will become the  best sample for BAO studies beyond redshift $z\ga 0.8$ until the arrival of BigBOSS at the end of this decade.
		However, this potential will not be realized without comprehensive redshift measurements.

		%-----------------------------------------------------------

		\subsection{Redshift data} \label{ssec:z}
		We assumed so far that the redshifts of all eRASS AGN are known.
		Now we briefly outline the requirements for the redshift data imposed by the science topics discussed above.
			
The linear bias as well as luminosity function studies do not demand a high accuracy of the redshift determination. % (Georgakakis et al. 2013, in prep.).
Indeed,  values of the order of $\delta z\sim 0.1-0.2$ should be sufficient, unless an analysis with a much higher redshift resolution is required.
In principle, this accuracy can be provided by photometric surveys.
However, one would need to investigate the impact of the large fraction of catastrophic errors, from which AGN redshift determinations based on the standard photometric filter sets are known to suffer \citep{Salvato2011}.
Of particular importance are redshift- and luminosity trends in catastrophic errors.
These problems will be considered in the forthcoming paper \citep{Huetsi2013}.
Provided that they are properly addressed,   optical photometric surveys of a moderate depth of $I\ga 22.5\,\mathrm{mag}$ \citep{Kolodzig2012} and with a sky coverage exceeding $\ga 2\,500\,\mathrm{deg^2}$ (Fig.~\ref{fig:SN_Amp_a}) would already produce first significant results.
		An existing survey with such parameters is SDSS.
		Its depth would allow detection of $\approx 80\%$ of eRASS AGN \citep{Kolodzig2012} and  with its sky coverage of $\sim14\,500\,\mathrm{deg^2}$  one should be able to conduct  high-accuracy measurements of the linear bias factor.
		Of the other ongoing surveys, the Pan-STARRS PS1 $3\pi$ survey \citep{Chambers2006} fulfills the necessary depth and sky coverage criteria.

		BAO studies, on the other hand, require a much higher redshift accuracy of the order of $\delta z\sim 0.01$.
		this accuracy can be only achieved in spectroscopic surveys or in high-quality  narrow-band multifilter photometric surveys.
		For example, for a $4\sigma$ detection of BAOs in the redshift range $0.8<z<1.2$, a spectroscopic survey of the depth of $I>22.5\,\mathrm{mag}$ is needed \citep{Kolodzig2012} and sky coverage of at least $\sim20\,000\,\mathrm{deg^2}$ (Fig.~\ref{fig:BAOs_Sigma}).
		Promising candidates are  the proposed 4MOST \citep{deJong2012} and WEAVE \citep{WEAVE} surveys, which would cover a large part of the sky with a multiobject spectrograph in the southern and northern hemisphere, respectively.
		An important caveat is that the angular resolution of eRASS (FOV averaged HEW of $\approx30^{\prime\prime}$) is insufficient, in particular for faint X-ray sources, to provide accurate sky positions for spectroscopic follow-ups with multiobject spectrographs.
		Therefore additional photometric surveys (for example Pan-STARRS PS1 $3\pi$) will be needed to refine source locations to the required accuracy.
		
		%-----------------------------------------------------------
		% Figure: BAO vs. Delta Z
		%-----------------------------------------------------------
		
			\begin{figure} %[htp]
				\resizebox{\hsize}{!}{\includegraphics{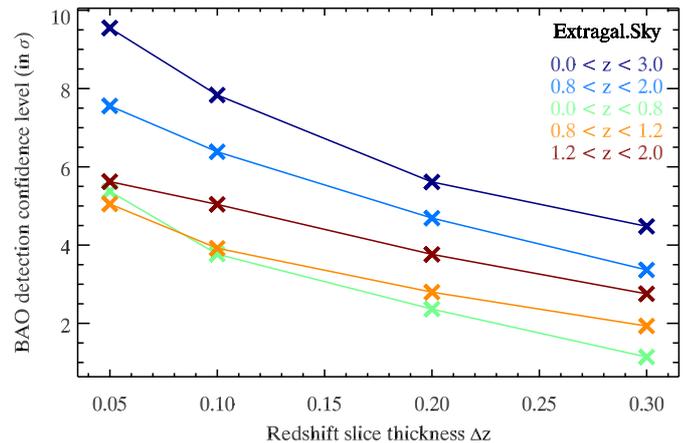}}   
				\caption{\label{fig:BAO_DeltaZ}%
					Confidence level of the BAO detection in the full extragalactic sky as a function of the redshift slice thickness (see Sect.~\ref{ssec:BAO.method}) for different redshift ranges. 
				}
			\end{figure}
		%-----------------------------------------------------------

	Fig.~\ref{fig:BAO_DeltaZ} shows the dependence of the confidence level of the BAO detection on the width of the redshift slice (Sect.~\ref{ssec:BAO.method}).
	This plot roughly illustrates how the accuracy of the BAO measurement deteriorates with decreasing accuracy of the redshift measurements.
	Although the confidence level clearly degrades by a factor of $\sim 2-3$, the BAOs should still be (marginally) detectable even with fairly high values of $\Delta z\sim 0.2-0.3$, characteristic for errors of the photometric redshifts obtained with a standard set of broad photometric filters \citep{Salvato2011}.
	However, as we discussed  earlier in this section, one of the significant problems of photometric redshifts is the large fraction of catastrophic errors.
	This factor was not accounted for in Fig.~\ref{fig:BAO_DeltaZ}.
	This problem will be addressed in full detail in the forthcoming paper by \citet{Huetsi2013}, which will consider BAO data analysis in a more general context, including a realistic simulation of photometric redshift errors and cross-terms in the power spectra.

		Finally, we note that we excluded from consideration  a number of observational effects and factors, such as source confusion and source detection incompleteness, positional accuracy, telescope vignetting, and nonuniformities in the survey exposure, as well as several others.
		These factors and effects  are well-known in X-ray astronomy, and data analysis methods and techniques  exist  to properly address them in the course of data reduction.

%		----------------------------------------------------------------------------------------------------------------------		
% 		\subsection{Redshift surveys}
% 		Where do we get our redshift information from? 
% 		
% 		\begin{description}
% 		 \item[Current]: \\
% 		 	\begin{itemize}
% 		        	\item SDSS(phot/spec) \citep{SDSSDR8,SDSSDR9}
% 		        	\item Pan-STARRS(phot) \citep{Chambers2006}
% 		        \end{itemize}
% 		 \item[Proposal]: \\
% 		 	\begin{itemize}
% 		        	\item eBOSS (phot)  (Kneib+2011, priv.\ comm. from A.Merloni)
% 		        	\item 4MOST (spec, southern, Europe) \citep{deJong2012}
% 		        	\item WEAVE (spec, nothern, USA) \citep{WEAVE}
% 		        	\item LSST(phot) \citep{LSST}
% 		        	\item Euclid (phot/spec) \citep{EUCLID}
% 		        \end{itemize}
%  		\end{description}
% 		List not complete! See also list in SB of eROSITA.
		
		%----------------------------------------------------------------------------------------------------------------------
		\section{Summary}
		%----------------------------------------------------------------------------------------------------------------------
		
% 		\subsection{Summary of results}
		We have explored the potential of the eROSITA all-sky survey for large-scale structure studies and  
		have shown that eRASS with its $\sim3$~million AGN sample  will supply us with outstanding opportunities for detailed LSS research.
		Our results are based on our previous work \citep{Kolodzig2012}, where we investigated statistical properties of AGN in eRASS, and on the AGN clustering model of  \citet{Huetsi2012}.
		
		We demonstrated that the linear bias factor of AGN can be studied with eRASS to unprecedented accuracy and detail. 
		Its redshift evolution can be investigated  with an accuracy of better than $\sim10\,\%$  using data from the sky patches of $\sim2\,500\,\mathrm{deg^2}$.
		Using the data from a sky area of  $\ga10\,000\,\mathrm{deg^2}$, statistically accurate redshift- and luminosity-resolved studies will become possible for the first time. Bias factor studies will yield meaningful results long before the full four-year survey will be completed.
		The eRASS AGN sample will not only improve the redshift- and luminosity resolution of bias studies but will also expand their luminosity range beyond $L_{0.5-2.0\,\mathrm{keV}}\sim10^{44}\,\mathrm{erg\;s^{-1}}$, thus enabling a direct comparison of the clustering properties of luminous X-ray AGN and optical quasars.
		These studies will dramatically improve our understanding of the AGN environment,  triggering mechanisms,  growth of supermassive black holes and their co-evolution with dark matter halos.
		The photometric redshift accuracy is expected to be sufficient for the bias factor studies, although the impact of the large fraction of catastrophic errors typical for standard broad-band filter sets needs yet to be investigated \citep{Huetsi2013}.
		
		For the first time for X-ray selected AGN, eRASS will be able to detect BAO with high-statistical significance of $\sim10\sigma$.
		Moreover, it will push the redshift limit of BAO detections far beyond the current limit of $z\sim0.8$.
		The accuracy of the BAO investigation in this uncharted redshift range will exceed that to be achieved by eBOSS, which is planned in the same timeframe, and will only be superseded  by BigBOSS, proposed for implementation after 2020.
		Until then, eRASS AGN can potentially become the best sample for BAO studies beyond $z\ga0.8$.
		However, for this potential to be realized and exploited, spectroscopic quality redshifts for large areas of the sky are required.

	%===========================================================
	% Appendix
	%===========================================================
		
		% Acknowledgements
		%-----------------------------------------------------------
 		\begin{acknowledgements} 
  			We thank Mirko Krumpe, Viola Allevato, Andrea Merloni, and Antonis Georgakakis for useful discussions and Chi-Ting Chiang for discussions of the HETDEX strategy and expected results.
  			
  			A. Kolodzig acknowledges support from and participation in the International Max-Planck Research School (IMPRS) on Astrophysics at the Ludwig-Maximilians University of Munich (LMU).
 		\end{acknowledgements} 
		
		% Bibliography
		%-----------------------------------------------------------
			\bibliographystyle{aa} 
			%\bibliography{references}

	%===========================================================
	% END of document
	%===========================================================
\end{document}